\documentclass[]{pasj02_orcid} 

\usepackage[dvipsnames]{xcolor}
\usepackage[switch,mathlines]{lineno} % add line number to manuscript
\usepackage{natbib} 
\definecolor{xlinkcolor}{cmyk}{1,1,0,0}
\usepackage{url}
\usepackage{comment}
\usepackage{lscape}
\usepackage{multirow}
\usepackage{threeparttable}
\usepackage{bm}

\usepackage[colorlinks=true, linkcolor=xlinkcolor, citecolor=xlinkcolor, urlcolor=xlinkcolor]{hyperref}
\makeatletter
\providecommand{\theH@article}{}

\makeatother

\jyear{2026}
\Received{}%{yyyy/mm/dd}
\Accepted{}%{yyyy/mm/dd}
\begin{document} 

\title{Merger-driven buildup of the $M_{\rm BH}\,\mathchar`-\,M_*$ relation bridging \hbox{high-$z$} overmassive black holes with the local relation}

\author{
Takumi S. \textsc{Tanaka},\altaffilmark{1,2,3} \email{takumi.tanaka@ipmu.jp} \orcid{0009-0003-4742-7060}
John D. \textsc{Silverman},\altaffilmark{1,2,3,4} \orcid{0000-0002-0000-6977}
Kazuhiro \textsc{Shimasaku},\altaffilmark{2,5} \orcid{0000-0002-2597-2231}
Knud \textsc{Jahnke},\altaffilmark{6} \orcid{0000-0003-3804-2137}
\\
Junyao \textsc{Li},\altaffilmark{7} \orcid{0000-0002-1605-915X}
Makoto \textsc{Ando}\altaffilmark{8,9} \orcid{0000-0002-4225-4477}
}
\altaffiltext{1}{Kavli Institute for the Physics and Mathematics of the Universe (WPI), The University of Tokyo Institutes for Advanced Study, The University of Tokyo, Kashiwa, Chiba 277-8583, Japan}
\altaffiltext{2}{Department of Astronomy, Graduate School of Science, The University of Tokyo, 7-3-1 Hongo, Bunkyo-ku, Tokyo 113-0033, Japan}
\altaffiltext{3}{Center for Data-Driven Discovery, Kavli IPMU (WPI), UTIAS, The University of Tokyo, Kashiwa, Chiba 277-8583, Japan}
\altaffiltext{4}{Center for Astrophysical Sciences, Department of Physics and Astronomy, Johns Hopkins University, Baltimore, MD 21218, USA}
\altaffiltext{5}{Research Center for the Early Universe, Graduate School of Science, The University of Tokyo, 7-3-1 Hongo, Bunkyo-ku, Tokyo 113-0033, Japan}
\altaffiltext{6}{Max Planck Institute for Astronomy, Königstuhl 17, 69117 Heidelberg, Germany}
\altaffiltext{7}{Department of Astronomy, University of Illinois at Urbana-Champaign, Urbana, IL 61801, USA}
\altaffiltext{8}{National Astronomical Observatory of Japan, 2-21-1 Osawa, Mitaka, Tokyo, 181-8588, Japan}
\altaffiltext{9}{Institute for Cosmic Ray Research, The University of Tokyo, 5-1-5 Kashiwanoha, Kashiwa, Chiba 277-8582, Japan}

%\footnotetext[$\dag$]{Present address: ....}

%%% end:list of authors

%% !!! Select 3 to 5 words from PASJ's key words !!! 
%% List of Key Words: https://academic.oup.com/pasj/pages/Pasj_Keywords 
%% "\KeyWords{ }" always has to be placed before ``\maketitle'' 

\maketitle
\begin{abstract}
The origin of the mass scaling relation between supermassive black holes (SMBHs, $M_{\rm BH}$) and galaxies ($M_*$) remains a key open question.
Rather than invoking AGN feedback, a non-causal mechanism has been proposed in which multiple mergers average out the $M_{\rm BH}/M_*$ ratio, thus decreasing its scatter ($\sigma$) and forming a tight local mass relation over cosmic history.
A larger scatter in the relation at higher redshift suggested from a non-causal evolutionary scenario may be evident from recent JWST observations of overmassive SMBHs at high redshift.
Here, we carry out a Monte Carlo simulation of solely merger-induced evolution of galaxies and their SMBHs which incorporates recent high-redshift observational constraints on $\sigma$ and the galaxy merger rate.
We find that the dispersion in the local mass relation can be reproduced, even when starting from a highly scattered population at $z\sim6$ with $\sigma=0.8\,{\rm dex}$ or $1.0\,{\rm dex}$, which are in agreement with recent JWST studies.
The redshift evolution of the scatter is highly sensitive to the mass ratio between merging pairs and the merger rate, and minor mergers with higher frequency than major mergers can also contribute to the scatter evolution, highlighting the importance of accurately constraining these parameters at high redshift through observations.
Furthermore, statistical surveys aimed at determining the $M_*$-dependence of $\sigma$ and constraining $\sigma$ at $z\sim3\,\mathchar`-\,4$ will be effective in testing this scenario. 
\end{abstract}
%\linenumbers
%\clearpage
% \pagewiselinenumbers

%%%%%%%%%%%%%%%%% Introduction %%%%%%%%%%%%%%%%%%
\section{Introduction}\label{s:intro}
In the local ($z\lesssim0.1$) Universe, a strong correlation is seen between the masses of galaxies (stellar mass, $M_*$) and their central supermassive black holes (SMBHs, $M_{\rm BH}$).
This local $M_{\rm BH}\,\mathchar`-\,M_*$ relation is typically expressed as a linear relation in logarithmic space:
\begin{equation}
    \log\left(\frac{M_{\rm BH}}{M_\odot}\right) = \alpha \log\left(\frac{M_*}{M_\odot}\right) + \beta, \label{eq:local}
\end{equation}
where $\alpha$ and $\beta$ are constants that vary depending on the particular population under study, with typical values of $\alpha\sim1$ (i.e., close to $M_{\rm BH}\propto M_*$) and $\beta\sim -2$ to $-3$.
The intrinsic scatter of this relation, i.e., scatter of $M_{\rm BH}$ from the equation (\ref{eq:local}), is reported to be around $\sigma \sim 0.2\,\mathchar`-\,0.4\,{\rm dex}$ \citep[e.g.,][]{Haring2004, Bennert2011, Kormendy2013, Reines2015}.
The process that forms this tight local relation remains a fundamental open question in our understanding of galaxy and SMBH evolution.

This tight relation may suggest that SMBHs and their host galaxies have co-evolved through direct physical coupling.
One possible mechanism is that active galactic nuclei (AGN) expel gas from the galaxy, depleting the fuel for both AGN and star formation activities \citep[AGN feedback, e.g.,][]{Springel2005, DiMatteo2008, Hopkins2008, Fabian2012, DeGraf2015, Harrison2017}.
Another causal mechanism is that the growth of SMBHs and galaxies has occurred in tandem, for example, triggered by the same phenomena like galaxy mergers \citep[e.g.,][]{Cen2015, Menci2016} or through torque-limited growth that enables balanced co-evolution of SMBHs and their host galaxies \citep{Angles2015, Angles2017}.
To verify these causal co-evolution scenarios, outflows in AGNs possibly contributing AGN feedback and a relation between star formation rates (SFR) and black hole accretion rates (BHAR) have been explored \citep[e.g.,][]{Yang2017_bhar, Fornasini2018, Carraro2020, Matsui2024}.
In addition, whether mergers truly trigger AGN activity has also been examined \citep{Ellison2011, Mechtley2016, Marian2019, Ellison2019}.
However, no definitive conclusion has yet been reached.

On the other hand, another important idea is that there may be no causal connection between SMBHs and galaxies; instead, repeated mergers statistically average out the $M_{\rm BH}/M_*$ ratio, eventually establishing the tight $M_{\rm BH}\,\mathchar`-\,M_*$ relation.
Through a central-limit-like process,  as time progresses and mergers accumulate, the $M_{\rm BH}/M_*$ ratio would naturally converge toward a constant value, and the $\sigma$ would decrease, naturally building the tight $M_{\rm BH}\,\mathchar`-\,M_*$ relation as an emergent statistical outcome \citep{Peng2007, Hirschmann2010, Jahnke2011}.
This idea has been tested through Monte Carlo simulations \citep{Peng2007, Hirschmann2010} and halo merger trees \citep{Jahnke2011}.
These studies found that the $\sigma$ indeed decreases with the number of mergers experienced.
In turn, this means that high-redshift populations with still fewer mergers should naturally exhibit a larger $\sigma$ than the local relation.
Therefore, examining the scatter of the $M_{\rm BH}\,\mathchar`-\,M_*$ relation at high redshift can test this non-causal evolution scenario.

The advent of the James Webb Space Telescope (JWST) has revolutionized the exploration of the high-$z$ $M_{\rm BH}\,\mathchar`-\,M_*$ relation.
JWST has successfully detected the host galaxies of high-redshift quasars, which were difficult to detect before JWST, initializing discussions on the relationship between quasars and their host galaxies at $z\sim6$ \citep[e.g.,][]{Ding2022_z6, Stone2023, Yue2023}.
Another notable population is ``Little Red Dots'' (LRDs), which are characterized by their compact morphologies, distinctive V-shaped SED from the rest-UV-to-optical wavelengths, and broad Balmer emission lines \citep[e.g.,][]{Labbe2023a, Harikane2023, Furtak2024, Kokorev2023, Matthee2024, Kocevski2023, Akins2023, Barro2024, Akins2024, Greene2024, Tanaka2025_z10}.
These studies reported higher $M_{\rm BH}/M_*$ ratios compared to the local $M_{\rm BH}\,\mathchar`-\,M_*$ relation when the $M_{\rm BH}$ were estimated using a single-epoch method with broad Balmer lines calibrated on low-redshift samples.

\cite{Li2024_iceburg} argue that the high-$z$ $M_{\rm BH}\,\mathchar`-\,M_*$ relation of the underlying population may be consistent with the local relation, showing no significant offset after considering the effects of a selection bias, which favors the detection of sources with large $M_{\rm BH}$.
Moreover, previous studies \citep{Li2025_moderate, Ren2025, Fei2025} identified high-redshift ($z>3$) systems with $M_{\rm BH}/M_*$ on (or below) the local relation, suggesting that the high-$z$ $M_{\rm BH}\,\mathchar`-\,M_*$ relation is not offset from the local relation.
Importantly, even in the case of no offset with the local relation, a larger intrinsic scatter than the local relation is still required to reproduce the observed overmassive population.
Despite large uncertainties due to the small samples, \cite{Li2024_iceburg} and \cite{Silverman2025} find $\sigma\sim0.97^{+0.52}_{-0.37}\,{\rm dex}$ at $z\gtrsim4$ and $\sigma\sim 0.80^{+0.23}_{-0.28}\,{\rm dex}$ at $z\sim6$, respectively.
Note that a recent study by \cite{Ziparo2026} also reported a scatter of $0.63^{+0.14}_{-0.11}\,{\rm dex}$, corresponding to a similar vertical (i.e., in the $M{\rm BH}$ direction) scatter of $\sigma \sim 0.96\,{\rm dex}$ at $z \sim 4\,\mathchar`-\,6$.
These larger scatters of the $M_{\rm BH}\,\mathchar`-\,M_*$ relation are possibly consistent with a merger-driven non-causal evolution.
This interpretation has also been supported by the discoveries of many AGNs in merging systems \citep[e.g.,][]{Harikane2023, Perna2023_highnumber, Ubler2024, Tanaka2024CB, Tanaka2024_dLRD}.

In this paper, we update Monte Carlo simulations of the merger-driven $\sigma$ evolution \citep{Peng2007, Hirschmann2010} with the latest observational constraints to test whether it is possible to reproduce the local tight relation with $\sigma\sim0.3\,{\rm dex}$ solely through mergers starting from $z\sim6$ population with large scatter inferred from JWST observations \citep{Li2024_iceburg, Silverman2025}.
In the simulations, we also use the redshift evolution of major merger rates constrained from observations, including JWST \citep{Duan2024}.
While SMBHs and galaxies also grow via secular processes such as gas accretion and star formation in the real Universe, we intentionally exclude such effects from our simulations to isolate the impact of mergers alone.
Uncertainties associated with assumptions about AGN feedback are therefore avoided in our simulations.
In sections\,\ref{s:simus} and \ref{s:results}, we describe the simulation setup and results, and in section\,\ref{s:discussion}, we discuss their implications and prospects for future studies.
Throughout this paper, we assume a standard cosmology with $H_0 = 70~{\rm km~s^{-1}~Mpc^{-1}}$, $\Omega_{\rm m} = 0.30$, and $\Omega_\Lambda=0.70$.

%%%%%%%%%%%%%%%%% METHODS %%%%%%%%%%%%%%%%%%
\section{Monte Carlo simulations}\label{s:simus}

In the simulations, we first generate initial values of $M_*$ and $M_{\rm BH}$ for each galaxy at $z=6$, according to the observed stellar mass function and the local $M_{\rm BH}\,\mathchar`-\,M_*$ relation (section\,\ref{ss:initial}).
We then select two galaxies from the mock sample and sum them at a redshift-dependent merger rate (section\,\ref{ss:merger_rate}) following two different strategies for galaxy replenishment (section\,\ref{ss:scenario}).
After repeating this process, we simulate the evolution of the $M_{\rm BH}\,\mathchar`-\,M_*$ relation under merger-driven growth.

The simulation proceeds in discrete time steps ($\Delta t$).
Since the merger rate $R_{\rm merger}$ is defined as the number of mergers per galaxy per unit time, the total number of merging galaxy pairs ($N_{\rm pair}$, i.e., the number of merger events) occurring during a single time step is given by
\begin{equation}
    N_{\rm pair} = \frac{1}{2} N_{\rm total}\left(t\right) R_{\rm merger}\left(t\right) \Delta t.
\end{equation}
where $N_{\rm total}\left(t\right)$ is the number of galaxies at time $t$.
In each step, $N_{\rm pair}$ pairs (i.e., $2N_{\rm pair}$ galaxies) that satisfy the stellar mass criteria described in section\,\ref{ss:merger_rate} are randomly selected from the galaxy sample to undergo a merger.
In each merger event, we assume that both $M_*$ and $M_{\rm BH}$ are simply added as
\begin{align*}
    M_{\rm BH, merge} &= M_{\rm BH,1} + M_{\rm BH,2},\\
    M_{\rm *, merge} &= M_{\rm *,1} + M_{\rm *,2},
\end{align*}
where $M_{\rm *,1}$, $M_{\rm *,2}$, $M_{\rm BH,1}$, and $M_{\rm BH,2}$ are the stellar masses and SMBH masses of each merging galaxy.
After each merger step, galaxies with $M_{\rm BH, merge}$ and $M_{\rm *, merge}$ are returned to the sample pool. 
Note that within the same step, no galaxy undergoes more than one merger.
To adopt finer time steps in the higher-$z$ Universe when the merger rate is higher, we divide the $\sim1.5\,{\rm Gyr}$ interval from $z=6$ to $z=2.7$ into 100 steps ($\Delta t\sim0.015\,{\rm Gyr}$), and the remaining period until $z = 0$ ($\sim11\,{\rm Gyr}$) into another 100 steps ($\Delta t\sim0.11\,{\rm Gyr}$).
We confirm that even when sampling the time steps 10 times more finely, the results remain unchanged.

\subsection{Initial distribution}\label{ss:initial}
For the initial distribution, we assume a population of low-$M_*$ SMBHs at $z \sim 6$, similar to those found in recent JWST observations.
First, we sample $M_*$ from the stellar mass function at $z=5.5\,\mathchar`-\,6.5$ fitted with a Schechter function as
\begin{align}
    \Phi\,{\rm d}\!\left(\log M_*\right) = &\ln\left(10\right) \Phi^* \exp\left(-10^{\log M_* - \log M^*}\right) \nonumber \\
    &\left(10^{\log M_* - \log M^*}\right)^{\alpha+1}\,{\rm d}\!\left(\log M_*\right),\label{eq:Schchter}
\end{align}
where $M^*=10^{10.93}M_\odot$, $\Phi=10^{-5.64}\,{\rm Mpc^{-3}\,dex^{-1}}$, and $\alpha=-2$ \citep{Shuntov2024}.
We limit the initial $\log\left(M_*/M_\odot\right)$ range into $\left[6.5,10\right]$. \cite{Hirschmann2010} tested both log-normal-like and Schechter-shaped initial distributions and found that the choice of distribution has minimal impact on the evolution of scatter. 
We adopt the results based on the Schechter distribution in this study.
We also confirm that adopting a log-normal-like distribution does not change the results.

\begin{figure}
 \begin{center}
  \includegraphics[width=8.5cm]{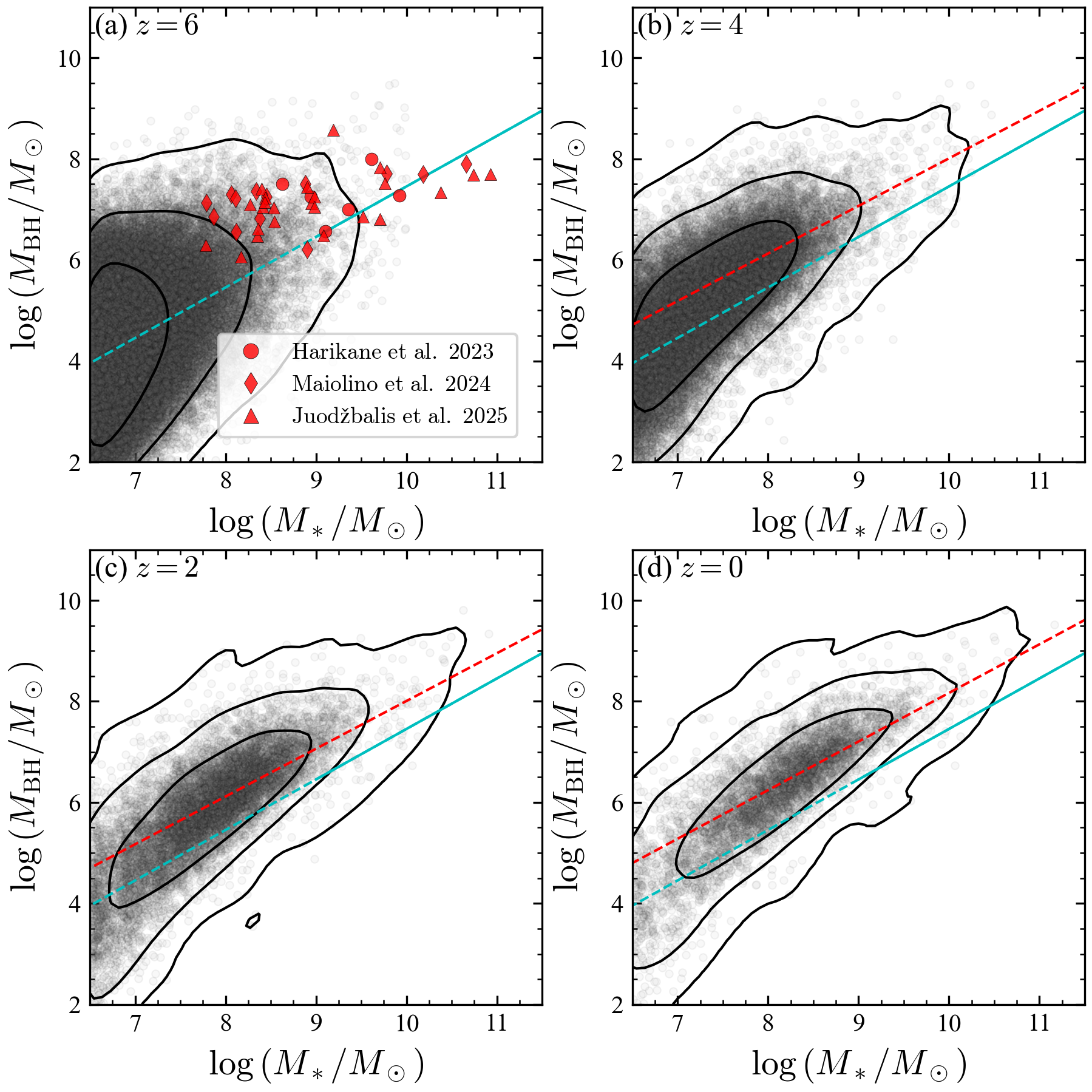} 
 \end{center}
\caption{Redshift evolution of the $M_{\rm BH}\,\mathchar`-\,M_*$ relation with $\sigma_{\rm init}=1\,{\rm dex}$ for case C ($\mu>1/40$) and the depletion scenario.
Panels (a)--(d) correspond to $z=6$ (initial distribution), $z=4$, $z=2$, and $z=0$.
Cyan and red lines indicate the local relation assumed to generate the initial distribution and the (log-)linear function (equation\,\ref{eq:local}) fitted to the objects with $\log M_*$ over their median at each redshift, respectively.
Black contours represent the distributions corresponding to the 68\%, 95\%, and 99\% distributions from inner to outer.
JWST-discovered high-$z$ AGNs \citep{Harikane2023, Maiolino2023, Joudzbalis2025} are shown in red circles, diamonds, and triangles in panel (a).
% \\Alt text: Four panels ($z=6$, $4$, $2$, and $0$) showing the simulated black hole mass versus stellar mass distributions.
% The x-axis shows the stellar mass, and the y-axis shows the black hole mass, both in units of the logarithm of solar mass.
}
\label{fig:mm}
\end{figure}

With mock stellar masses, we then assign values of $M_{\rm BH}$ to the sample by assuming the local relation with additional scatter following a normal distribution with the standard deviation of $\sigma_{\rm init}$ as
\begin{equation}
    \log\left(\frac{M_{\rm BH}}{M_\odot}\right) = \alpha \log\left(\frac{M_*}{M_\odot}\right) + \beta + \mathcal{N}\left(0,\sigma_{\rm init}\right), \label{eq:genarate_mbh}
\end{equation}
where we assume $\alpha=1$ and $\beta=-2.5$, following the local relation used in \cite{Tanaka2024} that is based on a local sample consisting of 30 inactive galaxies \citep{Haring2004} and 25 active galaxies \citep{Bennert2011}.
For $\sigma_{\rm init}$, we test two different values, $1.0\,{\rm dex}$ and $0.8\,{\rm dex}$, which are from \cite{Li2024_iceburg} and \cite{Silverman2025}, respectively.
Note that the differences may come from the mass range of the underlying populations: \cite{Li2024_iceburg} incorporates a wide range of sources, from low-luminosity AGNs (e.g., \citealt{Harikane2023} and \citealt{Maiolino2023}) to quasars (e.g., \citealt{Ding2022_z6}) with $\log\left(M_*/M_\odot\right)\sim7.5\,\mathchar`-\,11$ and $\log\left(M_{\rm BH}/M_\odot\right)\sim6.5\,\mathchar`-\,10$, while \cite{Silverman2025} use moderate-luminosity quasars from the SHELLQs sample at $z>6$ with $\log\left(M_*/M_\odot\right)\sim9.5\,\mathchar`-\,11$ and $\log\left(M_{\rm BH}/M_\odot\right)\sim8\,\mathchar`-\,9$.
Our initial distribution (see figure\,\ref{fig:mm} a) encompasses the observed properties of low-luminosity AGNs \citep{Harikane2023, Maiolino2023}.

\subsection{Merger rate}\label{ss:merger_rate}
\begin{figure*}
 \begin{center}
  \includegraphics[width=17cm]{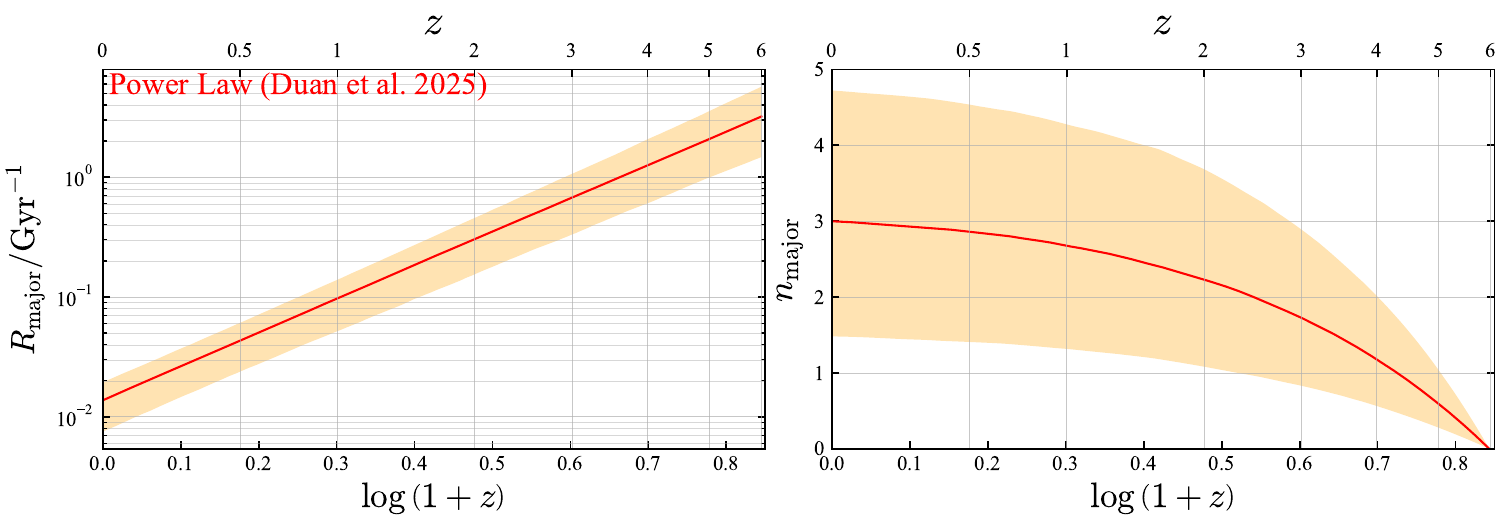} 
 \end{center}
\caption{
(Left) Major merger rate by \cite{Duan2024} as a function of redshift.
(Right) Cumulative number of major mergers ($n_{\rm major}$) from $z=6$ to each redshift.
In both panels, orange-shaded regions indicate 68\% confidence intervals.
% \\Alt text: Two-panel graphs.
% In the left panel, the x-axis shows the logarithm of one plus redshift, and the y-axis shows the merger rate in units of per gigayear.
% In the right panel, the x-axis shows the logarithm of one plus redshift, and the y-axis shows the cumulative number of major mergers.
}
\label{fig:merger_rate}
\end{figure*}

We define major mergers as mergers with $\mu >1/4$, where $\mu$ is the stellar mass ratio defined as
\begin{equation}
    \mu = \frac{M_{*,1}}{M_{*,2}}\hspace{7mm} (M_{*,1} < M_{*,2}),
\end{equation}
thus, $\mu\leq1$.
For the redshift dependence of the major merger rate ($R_{\rm major}\left(z\right)$, figure\,\ref{fig:merger_rate} left), we adopt the power law form presented by \cite{Duan2024}, which is based on observations over a wide redshift range
\begin{equation}
    R_{\rm major}\left(z\right) = \left(0.013\pm0.006\right) \times \left(1+z\right)^{\left(2.818\pm0.233\right)}.
\end{equation}
To evaluate how many major mergers occur from $z=6$, we define a major merger generation $n_{\rm major}\left(z\right)$ as an integration of the $R_{\rm major}\left(z\right)$ from $z=6$ to that redshift
\begin{equation}
    n_{\rm major}\left(z\right) = \int_z^6 R_{\rm major}\left(z^\prime\right) \left|\frac{{\rm d}t}{{\rm d}z^\prime}\right| {\rm d}z^\prime.\label{eq:nmajor}
\end{equation}
By definition, every galaxy will experience, on average, one major merger in $\Delta n_{\rm major}=1$, where $\Delta n_{\rm major}$ represents the change in $n_{\rm major}$ between two redshifts.
This means that if only major mergers are considered and we assume a closed simulation box without galaxy replenishment (depletion scenario introduced in section\,\ref{ss:scenario}), the number of galaxies decreases by a factor of $2^{-\Delta n_{\rm major}}$ over the interval of $\Delta n_{\rm major}$.
Figure\,\ref{fig:merger_rate}\,(right) suggests that, on average, each galaxy undergoes one major merger during each of the redshift intervals of $z=6\,\mathchar`-\,4$, $z=4\,\mathchar`-\,2$, and $z=2\,\mathchar`-\,0$.

As discussed in \cite{Peng2007} and \cite{Hirschmann2010}, minor mergers may also contribute to the formation of the tight $M_{\rm BH}\,\mathchar`-\,M_*$ relation.
However, due to observational limitations, minor merger rates at high redshift are not well constrained.
In the simulations, we derive the minor merger rate by simply scaling the major merger rate.
\cite{Rodriguez2015} derived the merger rate as a function of the descendant's mass, $\mu$, and $z$ (table\,1 in \citealt{Rodriguez2015}).
Following their equation, we roughly assume that mergers with $1/10 < \mu < 1/4$ (moderate mergers) occur at the same rate as major mergers, and those with $1/40 < \mu < 1/10$ (minor mergers) occur twice as frequently as major mergers.
To isolate the impact of different merger types, we compare three cases:\\
\noindent (A) only major mergers ($\mu>1/4$),\\
\noindent (B) major + moderate mergers ($\mu>1/10$),\\
\noindent (C) major + moderate + minor mergers ($\mu>1/40$).

\subsection{Depletion and replenishment scenarios}\label{ss:scenario}
In mock simulations of merger-driven $\sigma$ evolution, two approaches have been adopted \citep{Peng2007, Hirschmann2010}.
One is the depletion scenario, in which the initial population evolves within a closed box with no external addition, causing the number of galaxies to gradually decrease (see section\,\ref{ss:merger_rate}).
The other is the replenishment scenario, where new galaxies are added to offset the loss due to mergers.
Since galaxy formation continues in the real Universe, the replenishment scenario is considered the more realistic case.
\cite{Hirschmann2010} reported that the refill ratio, the number of newly added galaxies to the number of mergers, and the $\log M_*$ range of the added galaxies affect the resulting $\sigma$ evolution.

In this study, we implement both approaches and compare the results.
For the depletion scenario, as noted above, the galaxy counts decrease over time; therefore, the size of the initial distribution is set to 500,000 galaxies.
In contrast, for the replenishment scenario, the initial distribution contains 50,000 galaxies, and we assume that the number of galaxies replenished equals the number of mergers (i.e., refill-ratio is 1), thereby the sample size remains constant.
The replenished galaxies are generated in the same way as generating the initial distribution (section\,\ref{ss:initial}): we assume the same Schechter function (equation\,\ref{eq:Schchter}) for $M_*$ and generate $M_{\rm BH}$ following equation\,\ref{eq:genarate_mbh} with the same $\sigma_{\rm init}$.
We limit the $\log\left(M_*/M_\odot\right)$ of replenished galaxies to the range $\left[6, 8 \right]$, corresponding to the low-mass side of the initial distribution (figure\,\ref{fig:mm}), under the assumption that newly formed galaxies through structure formation are predominantly low-mass systems.
In sections\,\ref{ss:depletion} and \ref{ss:replenishment}, we present the results obtained with the depletion and replenishment scenarios, respectively.

%%%%%%%%%%%%%%%%% RESULTS %%%%%%%%%%%%%%%%%%
\section{Results}\label{s:results}

\subsection{Depletion scenario}\label{ss:depletion}

\subsubsection{Evolution on the $\bm{M_{\rm BH}\,\mathchar`-\,M_*}$ plane}\label{sss:dep:mbh-ms}
As shown in figure\,\ref{fig:mm}, mergers increase both $M_{\rm BH}$ and $M_*$ thus shifting the $M_{\rm BH}\,\mathchar`-\,M_*$ distribution toward the upper right direction on the $M_{\rm BH}\,\mathchar`-\,M_*$ plane as redshift decreases.
At the same time, the $M_{\rm BH}/M_*$ ratios become tighter due to averaging.
Note that in this merger-driven evolution scenario, galaxies grow toward the high-mass side, and the low-mass side remains largely unconstrained observationally.
Therefore, we fit a log-linear function for only galaxies with stellar masses greater than the median value, and the best-fit relation at each redshift is shown by red dashed lines in figure\,\ref{fig:mm}.
While the slope ($\alpha$) remains almost unchanged, the intercept ($\beta$) is higher than the initial relation (cyan line in figure\,\ref{fig:mm}), i.e., the final relation is considered overmassive.

One possible cause of this higher intercept is that $M_{\rm BH}$ is summed directly (i.e., not the sum of their logarithms) during each merger event, which biases the resulting sum toward higher $\log M_{\rm BH}$.
A similar effect could occur in the $M_*$ direction.
However, in this study, we limit the range of mass ratios ($\mu$), allowing $\mu > 1/40$ at most.
This restriction reduces the upward shift in $\log M_*$ relative to $\log M_{\rm BH}$, leading to a tendency to shift toward the overmassive side.
Because the resulting relation depends strongly on the assumptions for the initial distribution, which is highly uncertain at this stage, we do not focus on the absolute values of its slope and intercept.
Instead, we concentrate solely on the evolution of the scatter.

\begin{figure*}
 \begin{center}
  \includegraphics[width=16cm]{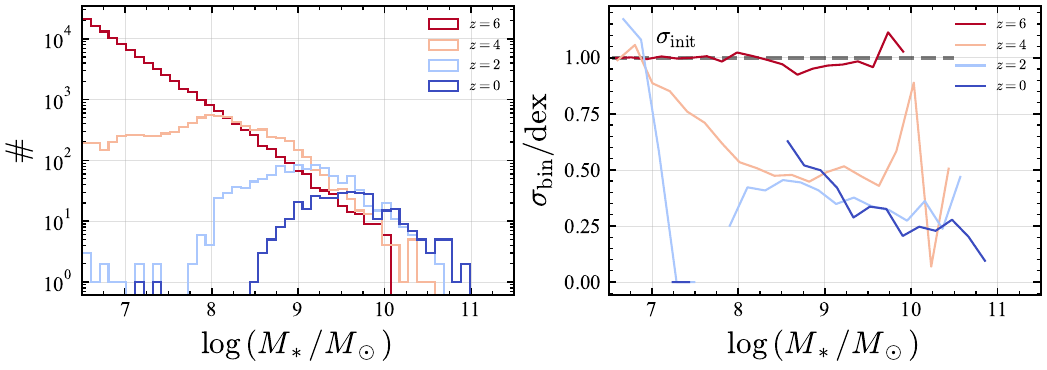} 
 \end{center}
\caption{
(Left) Distribution of $\log M_*$ at each redshift.
(Right) The $M_*$-dependence of the scatter of the $M_{\rm BH}\,\mathchar`-\,M_*$ relation in each $\log M_*$ bin ($\sigma_{\rm bin}$).
Both panels show the results for $\sigma_{\rm init}=1\,{\rm dex}$, case C, and the depletion scenario.
Line colors represent redshift, transitioning from red at $z=6$ (initial) to blue at $z=0$.
% \\Alt text: Two-panel graphs.
% The left panel is the histogram of the stellar mass in units of the logarithm of stellar mass.
% In the right panel, the x-axis shows the stellar mass in units of the logarithm of stellar mass, and the y-axis shows the scatter of the black hole mass at each stellar mass.
}
\label{fig:sigma_Ms}
\end{figure*}

The distribution of $\log M_*$ at each $z$ is shown in  figure\,\ref{fig:sigma_Ms}\,(left panel), which indicates that mergers shift the distributions to larger $M_*$.
On the high-mass side ($\log\left(M_*/M_\odot\right)\gtrsim 9.5$ at $z=0$), the $\log M_*$ distribution maintains the slope of the stellar mass function.
However, on the low-mass side, it deviates from a Schechter function, falling off more steeply due to the lack of supply from the lower-mass side in the depletion scenario.
This means that, on the lower-mass side where the Schechter function breaks down, the evolution of the population due to mergers may not be accurately reproduced because of the limited $M_*$ range of the initial sample.

\subsubsection{Evolution of the scatter}\label{sss:dep:sigma_evol}

\begin{figure*}
 \begin{center}
  \includegraphics[width=16cm]{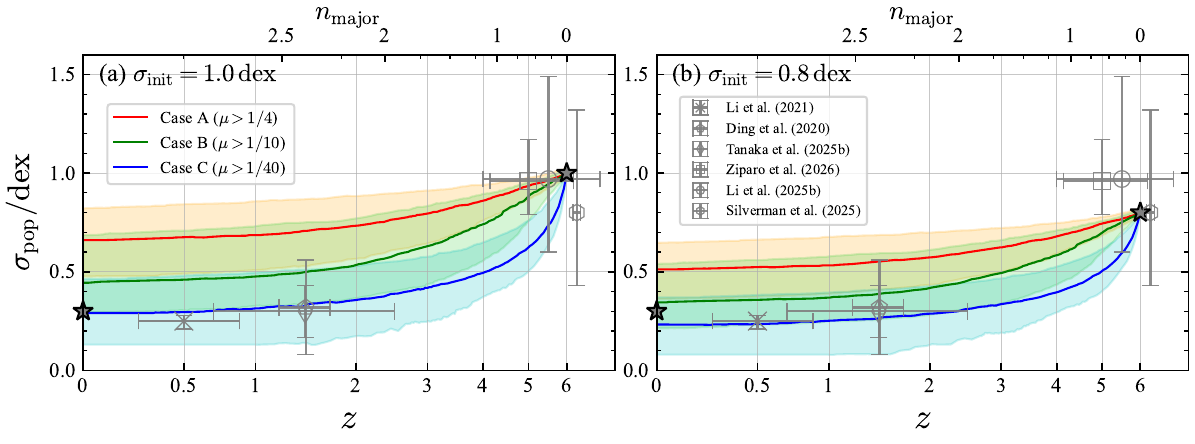} 
 \end{center}
\caption{
Redshift evolution of the scatter of the $M_{\rm BH}\,\mathchar`-\,M_*$ relation ($\sigma_{\rm pop}$) for the depletion scenario.
The left and right panels show the evolution starting from $\sigma_{\rm init}=1.0\,{\rm dex}$ and ${\rm 0.8\,{\rm dex}}$.
Red, green, and blue curves indicate the results for case A ($\mu>1/4$), case B ($\mu>1/10$), and case C ($\mu>1/40$).
Shaded regions indicate the 68\% confidence intervals.
Gray stars indicate the starting point of the simulations and the typical scatter of the local relation ($\sigma_{\rm pop}=0.3\,{\rm dex}$ at $z=0$).
Gray plots with error bars indicate the observational constraints on $\sigma$ \citep{Li2021_HSC, Ding2020_HST, Tanaka2024, Li2024_iceburg, Ziparo2026, Silverman2025}.
% \\Alt text: Two-panel graphs.
% In both panels, the x-axis shows the redshift, and the y-axis shows the scatter of the mass relation in units of dex.
}
\label{fig:sigma_z}
\end{figure*}

To estimate $\sigma$, we first fit the $M_{\rm BH}\,\mathchar`-\,M_*$ relation using a (log-)linear model (equation\,\ref{eq:local}).
We calculate the differences between the data and the fitted relation to evaluate $\sigma$ by fitting a Gaussian to the distribution of the residuals.
Figure\,\ref{fig:sigma_Ms}\,(right) shows $\sigma$ measured in each bin ($\sigma_{\rm bin}$) of $\log M_*$.
Since mergers increase $M_*$, higher-$M_*$ galaxies have smaller $\sigma_{\rm bin}$.
However, we note that this result may be influenced by the initial range in $\log\left(M_*/M_\odot\right)$ limited to $\left[6.5, 10\right]$.
In the depletion scenario, the low-mass population consists not only of galaxies that have grown through mergers, but also of galaxies that are directly added through replenishment.
Therefore, when estimating the scatter ($\sigma_{\rm pop}$) of the full population at each redshift, we fit a (log-)linear function (equation\,\ref{eq:local}) and evaluate $\sigma$ by fitting a Gaussian to the residuals using data above the first quartile in $M_*$.
In other words, we focus only on $\sigma$ for the high-mass side, which is accessible in observational studies.

The redshift evolution of $\sigma_{\rm pop}$ is shown in figure\,\ref{fig:sigma_z} for each merger scenario and two different starting values of $\sigma_{\rm init}$.
It is evident that $\sigma_{\rm pop}$ gradually decreases with redshift.
For an initial scatter of $\sigma_{\rm init} = 1.0\,{\rm dex}$, cases A and B only achieve scatters of 0.7\,dex and 0.5\,dex at $z = 0$, respectively.
On the other hand, case C reaches 0.3 dex, which is consistent with the observed scatter in the local and low‑z Universe \citep{Li2021_HSC, Tanaka2024}. 
When starting from $\sigma_{\rm init} = 0.8\,{\rm dex}$, case A attains a scatter of only 0.5\,dex, while case B, considering both major and moderate mergers, achieves 0.3\,dex, consistent with current local and low-$z$ observations.
These results demonstrate that the contributions from moderate and minor mergers are not negligible.
Interestingly, when case C is applied with $\sigma_{\rm init} = 0.8\,{\rm dex}$, the scatter at $z=0$ falls below 0.2\,dex, lower than the observed scatter.

\begin{figure}
 \begin{center}
  \includegraphics[width=8cm]{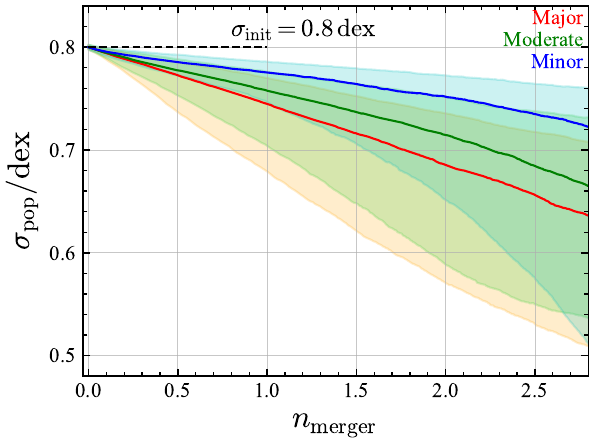} 
 \end{center}
\caption{Change in $\sigma_{\rm pop}$ as a function of the cumulative number of mergers from $\sigma_{\rm init}=0.8\,{\rm dex}$, assuming that each type of merger has the same merger rate.
Red, green, and blue lines correspond to major, moderate, and minor mergers.
Shaded regions indicate the 68\% confidence intervals.
% \\Alt text: Single panel figure.
% The x-axis shows the number of mergers, and the y-axis shows the scatter of the mass relation in units of dex, shown separately for major, moderate, and minor mergers.
}
\label{fig:sigma_z_compare}
\end{figure}

To exclusively compare the contributions of major ($1/4<\mu$), moderate ($1/10<\mu<1/4$), and minor mergers ($1/40<\mu<1/10$), we examine the evolution of $\sigma_{\rm pop}$ as a function of the average number of mergers ($n_{\rm merger}$) assuming that each type of merger occurs at the same rate.
Note that $n_{\rm merger}$ is similar to $n_{\rm major}$ in equation\,(\ref{eq:nmajor}), however, $n_{\rm merger}$ is defined for each type of merger.
Figure\,\ref{fig:sigma_z_compare} shows that mergers with lower mass ratios (i.e., moderate and minor mergers) contribute less to the decrease in $\sigma_{\rm pop}$ per merger event.
However, because we assume that minor mergers occur twice more frequently than major mergers based on simulation results \citep{Rodriguez2015}, the cumulative impact of minor mergers can exceed that of major mergers.
For example, the decrease in $\sigma_{\rm pop}$ from a single major merger ($n_{\rm merger}=1$) is comparable to that from approximately two minor mergers ($n_{\rm merger}=2$), demonstrating that repeated minor mergers can lead to a greater overall decrease in $\sigma_{\rm pop}$.

\subsection{Replenishment scenario}\label{ss:replenishment}
\begin{figure}
 \begin{center}
  \includegraphics[width=8.5cm]{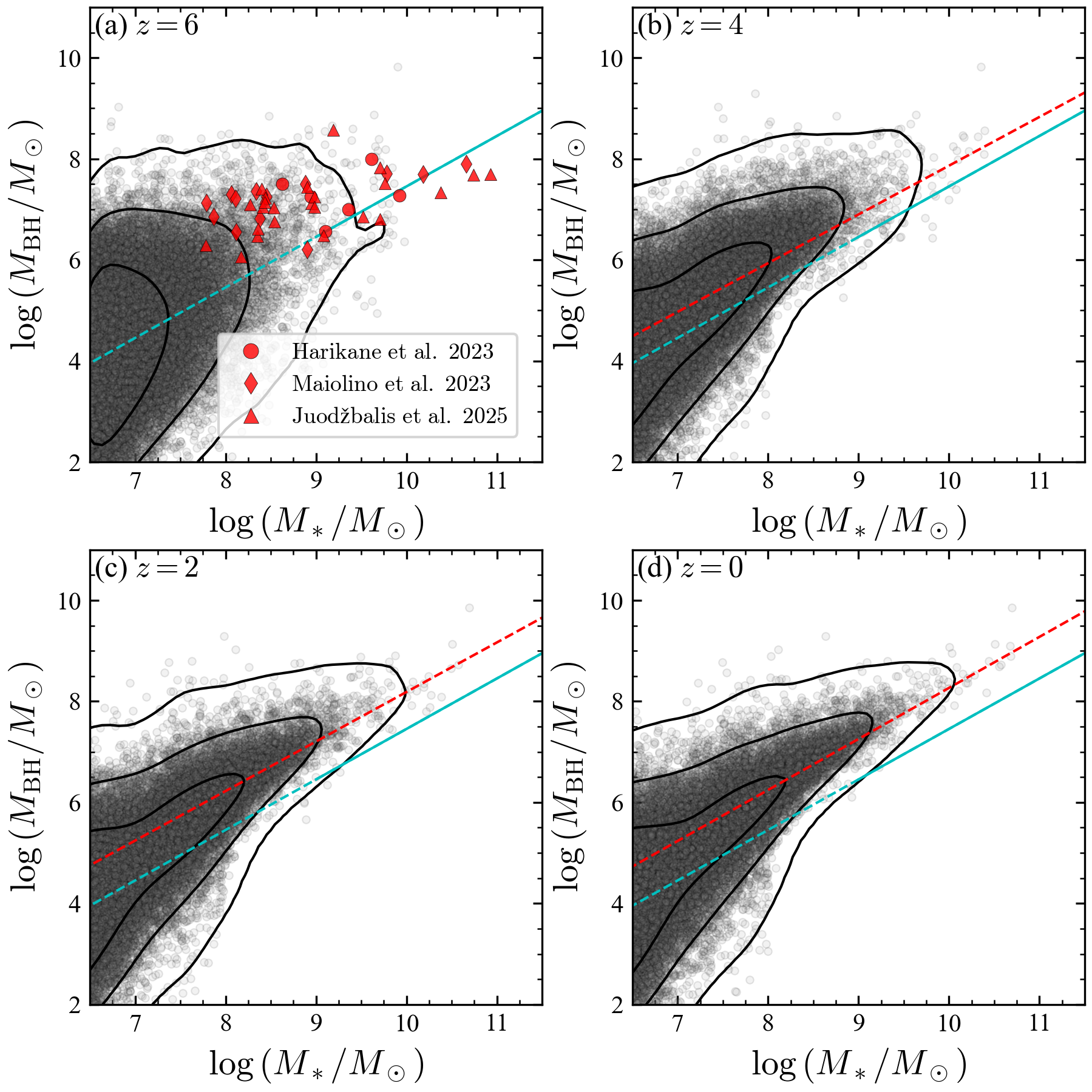} 
 \end{center}
\caption{
Same as in figure\,\ref{fig:mm}, but following the replenishment scenario.
% \\Alt text: Four panels ($z=6$, $4$, $2$, and $0$) showing the simulated black hole mass versus stellar mass distributions.
% The x-axis shows the stellar mass, and the y-axis shows the black hole mass, both in units of the logarithm of solar mass.
}
\label{fig:mm_re}
\end{figure}

\begin{figure*}
 \begin{center}
  \includegraphics[width=16cm]{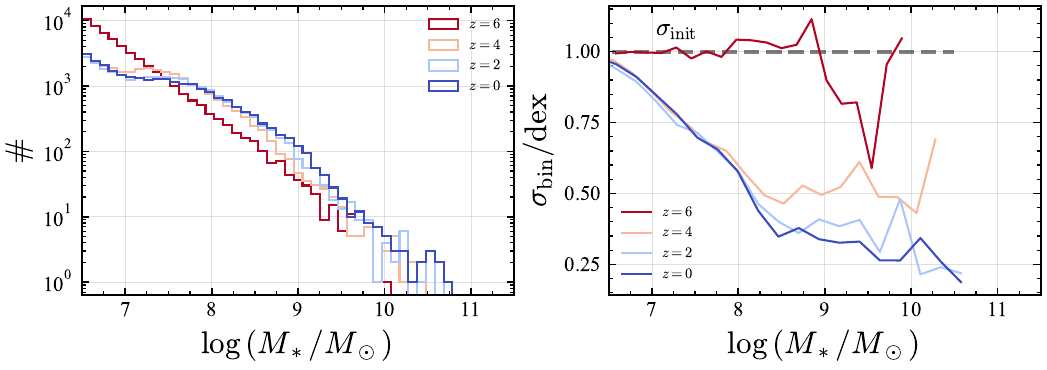} 
 \end{center}
\caption{
Same as in figure\,\ref{fig:sigma_Ms}, but following the replenishment scenario.
% \\Alt text: Two-panel graphs.
% The left panel is the histogram of the stellar mass in units of the logarithm of stellar mass.
% In the right panel, the x-axis shows the stellar mass in units of the logarithm of stellar mass, and the y-axis shows the scatter of the black hole mass at each stellar mass.
}
\label{fig:sigma_Ms_re}
\end{figure*}

Under the replenishment scenario (section\,\ref{ss:scenario}), new galaxies are replenished from the low-mass side following the Schechter function (section\,\ref{ss:scenario}, figure\,\ref{fig:mm_re}) to avoid the reduction of the overall number of galaxies through mergers. 
Unlike the depletion scenario, the $M_*$ distribution (figure\,\ref{fig:sigma_Ms_re}\,left panel) continues to follow the Schechter function down to $\log\left(M_*/M_\odot\right)\sim8$.
This enables the accurate modeling of minor mergers across a broader $M_*$ range.
On the $M_{\rm BH}\,\mathchar`-\,M_*$ plane (figure\,\ref{fig:mm_re}), the final $M_{\rm BH}\,\mathchar`-\,M_*$ relation is overmassive with a high $M_{\rm BH}/M_*$ ratio, similar to the results for the depletion scenario.

\begin{figure*}
 \begin{center}
  \includegraphics[width=16cm]{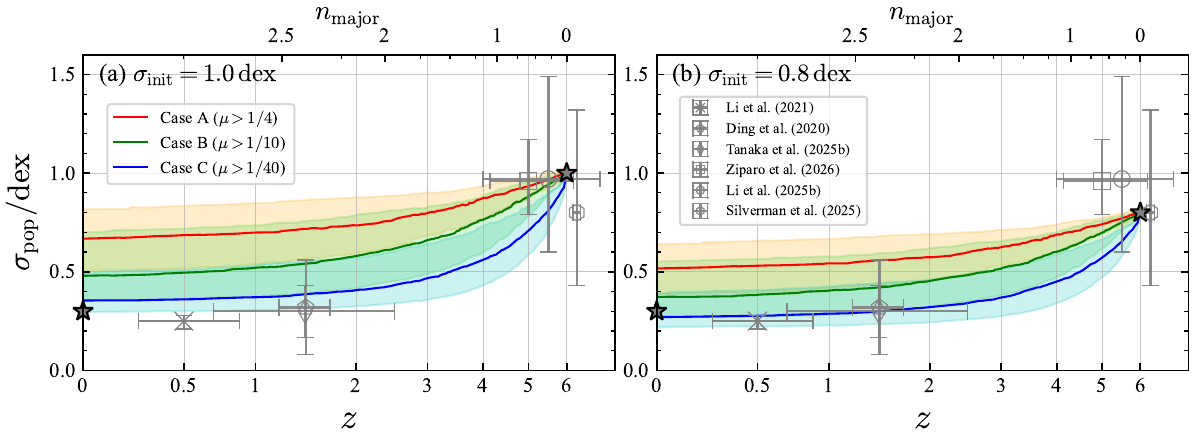} 
 \end{center}
\caption{
Same as in figure\,\ref{fig:sigma_z}, but following the replenishment scenario.
% \\Alt text: Two-panel graphs.
% In both panels, the x-axis shows the redshift, and the y-axis shows the scatter of the mass relation in units of dex.
}
\label{fig:sigma_z_re}
\end{figure*}

Because the low-mass side of the $M_*$ distribution is maintained in the replenishment scenario, we perform a (log-)linear fit (equation\,\ref{eq:local}) for objects with $\log\left(M_*/M_\odot\right)>8.5$.
Based on this fit, we estimate $\sigma_{\rm pop}$ by fitting a normal distribution to the residuals.
The redshift evolution of $\sigma_{\rm pop}$ (figure\,\ref{fig:sigma_z_re}) is similar to that of the depletion scenario.
As redshift decreases and mergers accumulate, $\sigma_{\rm pop}$ correspondingly decreases.
For $\sigma_{\rm init}=1.0\,{\rm dex}$ and $0.8\,{\rm dex}$, cases C and B, respectively, achieve a scatter of approximately $0.3\,{\rm dex}$ at $z=0$, which is consistent with local observational constraints.
These results indicate that even when starting with a large scatter of $\sigma_{\rm init}=1.0\,{\rm dex}$, the merger-driven evolution alone can reproduce the local tight mass relation with $\sigma\sim0.3\,{\rm dex}$ on the high-mass side until $z=0$ regardless of adopting the depletion or replenishment scenario.

%%%%%%%%%%%%%%%%% Discussion %%%%%%%%%%%%%%%%%%
\section{Discussion}\label{s:discussion}

\subsection{Implications of our simulations}
Our simulations suggest that the large scatter at high redshift can decline rapidly through mergers.
The observed abundance of overmassive systems at high redshift relative to low redshift may arise not only due to selection bias \citep{Li2024_iceburg} but also because such overmassive objects rapidly disappear through mergers and become rarer by $z\sim2$.
If this merger-driven evolutionary scenario is correct, then the $M_{\rm BH}\,\mathchar`-\,M_*$ relation is not the result of a one-sided ``SMBH-first'' or ``galaxy-first'' process but rather arises from the averaging effect of populations that have evolved randomly and then merged.

Our simulations do not include secular processes.
Therefore, especially in the depletion scenario where no new galaxies are added, $M_*$ of our simulations should trace the older stellar populations.
In this sense, the mass relation reproduced especially at lower-$z$ may roughly correspond to the $M_{\rm BH}\mathchar`-M_{\rm bulge}$ relation in the real Universe.
However, note that we do not account for the $M_*$ loss after stellar lifetime, nor do we solve the dynamics that separate bulges and disks within each galaxy; thus, a fully rigorous discussion is not possible.

Furthermore, our simulations produce an overmassive relation at $z=0$, i.e., a larger intercept $\beta$ than the local relation.
Since the final distribution depends on the initial conditions, explaining the evolution solely via mergers requires that the relationship in the early Universe is not centered on the local relation.
Instead, the initial distribution may have a lower intercept ($\beta$) than that of the local relation, or it could exhibit a mass function that is more strongly weighted toward lower $M_{\rm BH}$ values, i.e., not a log-normal distribution.

Previous theoretical studies using cosmological hydrodynamic simulations and semi-analytic models have explored the redshift evolution of the $M_{\rm BH}\,\mathchar`-\,M_*$ (or $M_{\rm BH}\,\mathchar`-\,M_{\rm bulge}$) relation, including its scatter.
In the $\nu^2$GC semi-analytic model, \cite{Shimizu2024} reported that, at $z\gtrsim3$, two sequences emerge in the $M_{\rm BH}\,\mathchar`-\,M_{\rm bulge}$ plane, which they attribute to different dominant growth triggers of galaxy mergers and disc instabilities.
In their model, merger-driven growth produces a high-$M_{\rm BH}$ sequence, and at lower redshifts, increasingly gas-poor, dry mergers preferentially grow bulges and help bridge the two sequences, leading to convergence toward a single local relation.
While our merger-only experiment is not designed to reproduce the full multi-channel behavior seen in such models with secular processes, our simulations instead provide a reference case that isolates the purely statistical ``averaging'' effect of mergers.
Both our simulations and \cite{Shimizu2024} similarly suggest that repeated mergers can reduce the intrinsic scatter and contribute to the formation of tight mass relations at lower redshifts.

\subsection{Prospects}
In our simulations, we only include mergers with a mass ratio of $\mu > 1/40$.
Although individual minor mergers with $1/40<\mu<1/10$ have a smaller effect on decreasing $\sigma$ than major mergers with $1/4<\mu<1$ (figure\,\ref{fig:sigma_z_compare}), their higher frequency leads to a cumulative non-negligible contribution (figures\,\ref{fig:sigma_z} and \ref{fig:sigma_z_re}).
Therefore, incorporating mergers with even lower $\mu$ values ($\mu<1/40$) might lead to a larger decrease in $\sigma_{\rm pop}$.
However, including mergers with smaller $\mu$ would require constructing a sample that spans a wider range in $M_*$, and thus, a larger initial sample requesting greater computational cost.
In addition, although our study distinguishes between only three types of mergers (major, moderate, and minor), subdividing the $\mu$ range more finely could allow us to evaluate each merger's impact on the overall evolution accurately.

To improve the accuracy of the simulations, it is important to impose tighter observational constraints on the merger rate at high redshift.
Moreover, simulation studies suggest that the merger rate also depends on $M_*$.
For simplicity, our simulation assumes a uniform merger rate independent of $M_*$.
In future work, incorporating $M_*$-dependence as proposed by \cite{Rodriguez2015} would enable a more realistic representation of the underlying processes.
To account for these effects, including the environmental dependence of mergers, we also plan to revisit and update merger-tree-based simulations \citep{Jahnke2011} with the recent high-$z$ observational constraints.

In SMBH mergers, $\lesssim10\%$ of the mass is thought to be released as gravitational waves.
For example, after roughly three major mergers ($n_{\rm major}=3$), the combined mass might be reduced to about 70\% of its initial value.
If this fractional mass loss is constant regardless of the absolute mass, then in logarithmic space, it simply results in a constant vertical shift without affecting the $\sigma$ evolution.
A reduction to 70\% corresponds to $\sim-0.15\,{\rm dex}$ shift in the $\log M_{\rm BH}$, possibly affecting the final $M_{\rm BH}\,\mathchar`-\,M_*$ relation.

It is also important to note that our simulations do not include secular processes to isolate the baseline, statistical effect of merger averaging on the evolution of the intrinsic scatter, without introducing additional model dependencies associated with uncertain prescriptions for baryonic physics.
While this study demonstrates that mergers alone can produce a tight $M_{\rm BH}\,\mathchar`-\,M_*$ relation without any balanced secular processes (e.g., feedback), our results do not imply that secular processes are unimportant in the mass relation formation in the real Universe.
In fact, cosmological hydrodynamic simulations show that the normalization, shape, and scatter of the $M_{\rm BH}\,\mathchar`-\,M_*$ relation can differ substantially among models, underscoring the potential impact of BH accretion and feedback implementations on the mass relation evolution \citep[e.g.,][]{Habouzit2021, Habouzit2022}.
Furthermore, if mergers and secular processes are correlated, our simulations would not accurately capture the merger contribution.
One of the challenges in incorporating secular processes is that we need some assumptions about the star formation histories (SFHs) and black hole accretion histories (BHAHs).
For future work, it would be necessary to incorporate secular processes by exploring a range of scenarios--from simple SFH and BHAH assumptions to the most extreme cases like bursty history--and investigate their effects on merger-induced evolution.

Our simulations predict a $\sigma_{\rm pop}\sim0.4\,\mathchar`-\,0.6\,{\rm dex}$ at $z \sim 3\,\mathchar`-\,4$.
Thus, performing a statistical survey at $z \sim 3\,\mathchar`-\,4$ and constraining the $\sigma_{\rm pop}$ in this redshift range is crucial to test this scenario.
Another important observational task is to examine the $M_*$-dependence of $\sigma_{\rm bin}$. 
As shown in figure\,\ref{fig:sigma_Ms}, when the evolution is driven solely by mergers, $\sigma_{\rm bin}$ is expected to decrease as $M_*$ increases.
Although current observations are not yet sufficient to explore this dependence, expanding the sample through dedicated observational efforts will be a crucial next step.

\section{Conclusion}
Several studies have attempted to model the evolution of the scatter of the $M_{\rm BH}\,\mathchar`-\,M_*$ relation ($\sigma$) via mergers \citep{Peng2007, Hirschmann2010, Jahnke2011}.
We revisit the use of Monte Carlo simulations of the merger-driven $\sigma$ evolution and update them with the latest observational results.
Our simulations use the observational constraints on $\sigma$ at $z\sim6$ from JWST as the initial scatter \citep{Li2024_iceburg, Silverman2025} and major merger rates derived from observations \citep{Duan2024}.
In this way, our study may serve as a test closer to reality than previous studies.
In addition, we explicitly distinguish between major, moderate, and minor mergers while performing quantitative comparisons that account for differences in their occurrence rates.
The key results from this simulation are as follows:
\begin{itemize}
    \item Major mergers alone ($\mu>1/4$) are insufficient to explain the evolution of the mass ratio from large $\sigma$ inferred from JWST results to a tight relation with $\sim0.3\,{\rm dex}$ scatter at $z\sim0$.
    \item The inclusion of moderate ($1/10<\mu<1/4$) and minor mergers ($1/40<\mu<1/10$) contributes to decreasing $\sigma$ to $\sim0.3\,{\rm dex}$ at $z\sim0$ and explaining the observationally inferred redshift evolution of $\sigma$ without a further secular process.
\end{itemize}
However, this study does not rule out the possibility that the $M_{\rm BH}\,\mathchar`-\,M_*$ relation could also be established through secular processes, such as AGN feedback.
Further improvement of the model, including the incorporation of mergers with even lower mass ratios ($\mu < 1/40$), the effects of secular processes, star formation, and black hole accretion will be essential for a deeper understanding of the contribution of mergers to the emergence of the tight local $M_{\rm BH}\,\mathchar`-\,M_*$ relation.
Also, it is important to measure the scatter and the merger rate at higher redshifts more accurately.
In particular, the scatter at $z \sim 3\,\mathchar`-\,4$ serves as a key bridge between the frequently studied redshift ranges at $z \lesssim 2$ and $z \sim 6$, and therefore dedicated surveys aimed at constraining $\sigma$ in this intermediate range would be an important step.
It will also be helpful to examine the mass dependence of $\sigma$, as this is a critical observational step toward constraining the origin and evolution of the mass relation.

\begin{ack}
Numerical computations were in part carried out on the iDark cluster, Kavli IPMU.
We thank Daniel Angles-Alcazar for the fruitful discussion and Francesco Ziparo for sharing their results.
We thank the anonymous referee for helpful feedback.
\end{ack}

\section*{Funding}
Kavli IPMU is supported by World Premier International Research Center Initiative (WPI), MEXT, Japan.
TT is supported by Japan Society for the Promotion of Science (JSPS) KAKENHI Grant Number JP25KJ0750, a grant from the Hayakawa Satio Fund awarded by the Astronomical Society of Japan, and the Forefront Physics and Mathematics Program to Drive Transformation (FoPM), a World-leading Innovative Graduate Study (WINGS) Program at the University of Tokyo.
MA acknowledges that this work was supported by the Data-Scientist-Type Researcher Training Project of The Graduate University for Advanced Studies, SOKENDAI.

% \section*{Data availability} 
% The data underlying this article are available ...  
% Sample Data Availability Statements 
% https://academic.oup.com/pages/open-research/research-data#Data%20Availability%20Statements

% Any journal's BST file (e.g., apj.bst) can be used as PASJ's BST is unavailable.    
\bibliographystyle{plainnat2}
\bibliography{export}{}
% \bibliographystyle{****}
% \bibliography{****}

\end{document}